\documentclass[3p,times,procedia]{elsarticle}
\usepackage{nupha_ecrc}

\volume{00}

\firstpage{1}

\journalname{Nuclear Physics A}

\runauth{}

\jid{nupha}

\jnltitlelogo{Nuclear Physics A}

\usepackage{amssymb}

\usepackage[figuresright]{rotating}

\begin{document}

\begin{frontmatter}



\dochead{}

\title{Extracting $\hat{q}$ in event-by-event hydrodynamics and the centrality/energy puzzle}


\author[label1]{Carlota Andres}
\author[label1]{Nestor Armesto}
\author[label2]{Harri Niemi}
\author[label1,label3]{Risto Paatelainen}
\author[label1]{Carlos A. Salgado}
\author[label4]{Pia Zurita}

\address[label1]{Departamento de F\'isica de Part\'iculas and IGFAE, Universidade de Santiago de Compostela, 15782 Santiago de Compostela, Spain}
\address[label2]{Institut f\"ur Theoretische Physik, Johann Wolfgang Goethe-Universit\"at, Max-von-Laue-Str. 1, D-60438 Frankfurt am Main, Germany}
\address[label3]{University of Jyvaskyla, Department of Physics, P.O.B. 35, FI-40014 University of Jyvaskyla, Finland}
\address[label4]{Physics Department, Brookhaven National Laboratory, Upton, NY 11973, USA}

\begin{abstract}
In our analysis, we combine event-by-event hydrodynamics, within the EKRT formulation, with jet quenching -ASW Quenching Weights- to obtain high-$p_T$ $R_{\rm AA}$ for charged particles at RHIC and LHC energies for different centralities. By defining a $K$-factor that quantifies the departure of $\hat{q}$ from an ideal estimate, $K = \hat{q}/(2\epsilon^{3/4})$, we fit the single-inclusive experimental data for charged particles. This $K$-factor is larger at RHIC than at the LHC but, surprisingly, it is almost independent of the centrality of the collision.
\end{abstract}

\begin{keyword}
jet quenching \sep event-by-event hydrodynamics  \sep energy loss
\end{keyword}

\end{frontmatter}


\section{Introduction}
\label{sect:intro}
Jet quenching is a fruitful tool to extract medium parameters that characterize the quark-gluon plasma formed in high-energy nuclear collisions. We perform here an extraction of the $\hat{q}$ parameter using RHIC and LHC data on the nuclear modification factor, $R_{\rm AA}$, for single-inclusive particle production at high transverse momentum. The formalism of Quenching Weights \cite{Baier:2001yt,Salgado:2002cd,Salgado:2003gb}, embedded in EKRT event-by-event (EbyE) hydrodynamic model of the medium \cite{Niemi:2015qia}, is used.

We define the jet quenching parameter $K \equiv \hat{q}/(2\epsilon^{3/4})$, motivated by the ideal estimate $\hat q_{\rm ideal}\sim 2\epsilon^{3/4}$ \cite{Baier:2002tc}, where $\epsilon$ is the energy density given by the EKRT hydrodynamic description. Our main conclusions are that this $K$-factor is $\sim2 - 3$ times larger for RHIC than for the LHC and, unexpectedly, it is not dependent on the centrality of the collision.

\section{Jet quenching formalism}
\label{sect:jetquenching}
Our analysis is restricted to the simplest observable, the nuclear modification factor, $R_{AA}$, given by:
\begin{equation}
R_{AA} = \frac{dN_{AA}/d^2p_Tdy}{\langle N_{coll}\rangle dN_{pp}/dp_T^2dy};
\label{eqn1}
\end{equation}
hence, both the vacuum and the medium single-inclusive cross sections need to be calculated.

The cross section of a hadron $h$ at rapidity $y$ and transverse momentum $p_T$ can be described by
\begin{equation}
\frac{d\sigma^{AA\rightarrow h+X}}{dp_Tdy} = \int \frac{dx_2}{x_2}\frac{dz}{z}\sum\limits_{i,j}x_1f_{i/A}(x_1,Q^2)x_2f_{j/A}(x_2,Q^2)\frac{d\hat{\sigma}^{ij\rightarrow k}}{d\hat{t}}D_{k \rightarrow h}(z, \mu_F^2), 
\label{crosssection}
\end{equation}
where $A$ is the mass number of the nucleus, so $A=1$ for the vacuum cross section. $f_{i/A}(x_1,Q^2)$ are the PDFs, $d\hat{\sigma}^{ij\rightarrow k}/d\hat{t}$ the partonic cross section and $D_{k \rightarrow h}(z, \mu_F^2)$ the fragmentation functions.

All these computations are done at NLO using the code \cite{Stratmann:2001pb}, with the proton PDF set CTEQ6.6M \cite{Nadolsky:2008zw} and DSS vacuum fragmentation functions \cite{deFlorian:2007aj}. The renormalization, fragmentation and factorization scales are taken as $\mu_F = p_T$. For the medium cross section, EPS09 nPDFs \cite{,Eskola:2009uj} are used and the energy loss is absorbed in a redefinition of the fragmentation functions: 
\begin{equation}
D^{(med)}_{k\rightarrow h}(z,\mu_F^2)=\int \limits_0^1d\epsilon P_E(\epsilon)\frac{1}{1 - \epsilon}D^{(vac)}_{k\rightarrow h}\left(\frac{z}{1 - \epsilon},\mu_F^2\right)\ ,
\label{medfragmentations}
\end{equation}
where $P_E(\epsilon)$ are the ASW Quenching Weights.

The Quenching Weights are the probability distribution of a fractional energy loss, $\epsilon = \Delta E/E$, of the fast parton in the medium. They are based on two main assumptions: fragmentation functions are not medium-modified and gluon emissions are independent. These are good approximations for the total coherence case and for soft radiation  \cite{CasalderreySolana:2012ef,Blaizot:2012fh,Armesto:2007dt}. Indeed,  QW and rate equations are equivalent for soft radiation and no finite energy effects. In our study, the QW are used in the multiple soft approximation.

The quenching weights, $P_i(\Delta E/\omega_c,R)$, are dependent on two variables: $\omega_c=\frac{1}{2}\hat qL^2,$ and $R = \omega_cL$. These variables, can be obtained for a dynamic medium by \cite{Salgado:2002cd}
\begin{equation}
\omega_c^{eff}(x_0,y_0,\tau_{\rm prod},\phi)=\int d\xi\,\xi\,\hat q(\xi) , \qquad R^{eff}(x_0,y_0,\tau_{\rm prod},\phi)=\frac{3}{2}\int d\xi\,\xi^2\, \hat q(\xi).
\label{eq:omceff}
\end{equation}

So, we only need to specify the relation between the local value of the transport coefficient $\hat q(\xi)$ at a given point of the trajectory and  the hydrodynamic properties of the medium:
\begin{equation}
\hat q(\xi)=K\cdot 2\epsilon^{3/4}(\xi),
\label{eq:qhatlocal}
\end{equation}
where $K\simeq 1$ would correspond to the ideal QGP \cite{Baier:2002tc}. The local energy density $\epsilon(\xi)$ is taken from the EKRT simulations \cite{Niemi:2015qia}.

\section{EKRT hydrodynamics}
\label{sect:hydro}
We obtain the event-by-event space-time distribution of the local energy density by solving the relativistic hydrodynamic equations with EKRT initial state, with constant shear viscosity $\eta/s=0.2$ and starting time of viscous hydrodynamics $\tau_0 = 0.197$ fm \cite{Niemi:2015qia}. In our previous analysis several smooth-averaged hydrodynamic simulations were used \cite{Andres:2016iys}. We show here that our current results are compatible with the previous ones.

There is an ambiguity on the definition of $\hat{q}$ (\ref{eq:qhatlocal}) for times smaller than the thermalization time $\tau_0$. Nevertheless, as $\tau_0$ for the EKRT hydro is much smaller than for the smooth-averaged ones, the differences coming from the various extrapolations for times prior to thermalization are reduced. Hence, we consider here only one extrapolation:
\begin{equation}
\hat q(\xi)=\hat q(\tau_0) \quad \mathrm{for} \quad \xi<\tau_0.
\label{eqn:prior}
\end{equation}

\section{Results}
\begin{figure*}
\includegraphics[scale=0.35]{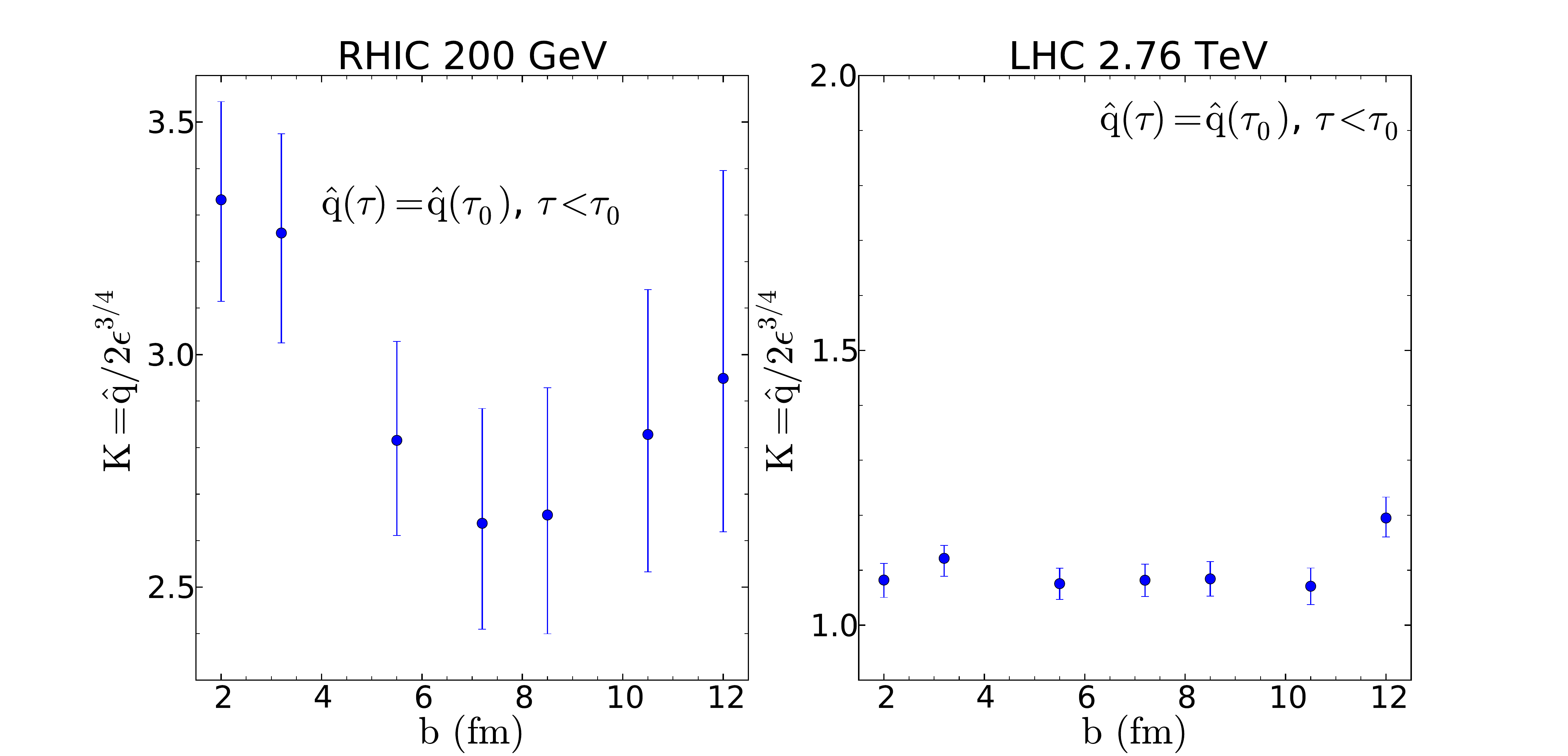}
\caption{$K$-factors obtained from fits to PHENIX $R_{\rm AA}$ data \cite{Adare:2008qa} \textit{(left panel)} and to ALICE $R_{\rm AA}$ data \cite{Abelev:2012hxa} \textit{(right panel)} versus the average impact parameter for each centrality class, $b$.}
\label{fig:chi2RHICandLHC}
\end{figure*}

We study the nuclear modification factor, $R_{\rm AA}$, both at RHIC \cite{Adare:2008qa} and the LHC \cite{Abelev:2012hxa} at different centralities. We have performed a $\chi^2$ fit to the best value of $K$ for each energy and centrality. The uncertainty band is determined by $\Delta \chi ^2 = 1$. In the left panel of Fig.~\ref{fig:chi2RHICandLHC}, we plot the different values of the K-parameter fitted to the PHENIX data \cite{Adare:2008qa}. The corresponding values for the LHC \cite{Abelev:2012hxa} are plotted in the right panels of the same figures.

First of all, the fitted value at RHIC confirms large corrections to the ideal case, while the corresponding one at the LHC is close to the unity. The $K$-factor obtained is $\sim2 - 3$ times larger for RHIC than for the LHC. Other groups \cite{Burke:2013yra} have found a factor $\sim 1.25$. Second, the LHC results are constant except for the most peripheral collisions. Consequently, the fitted value of $K$ seems to be primarily dependent on the energy of the collision and not to depend on the centrality of the collision.

\begin{figure*}
\includegraphics[scale=0.325]{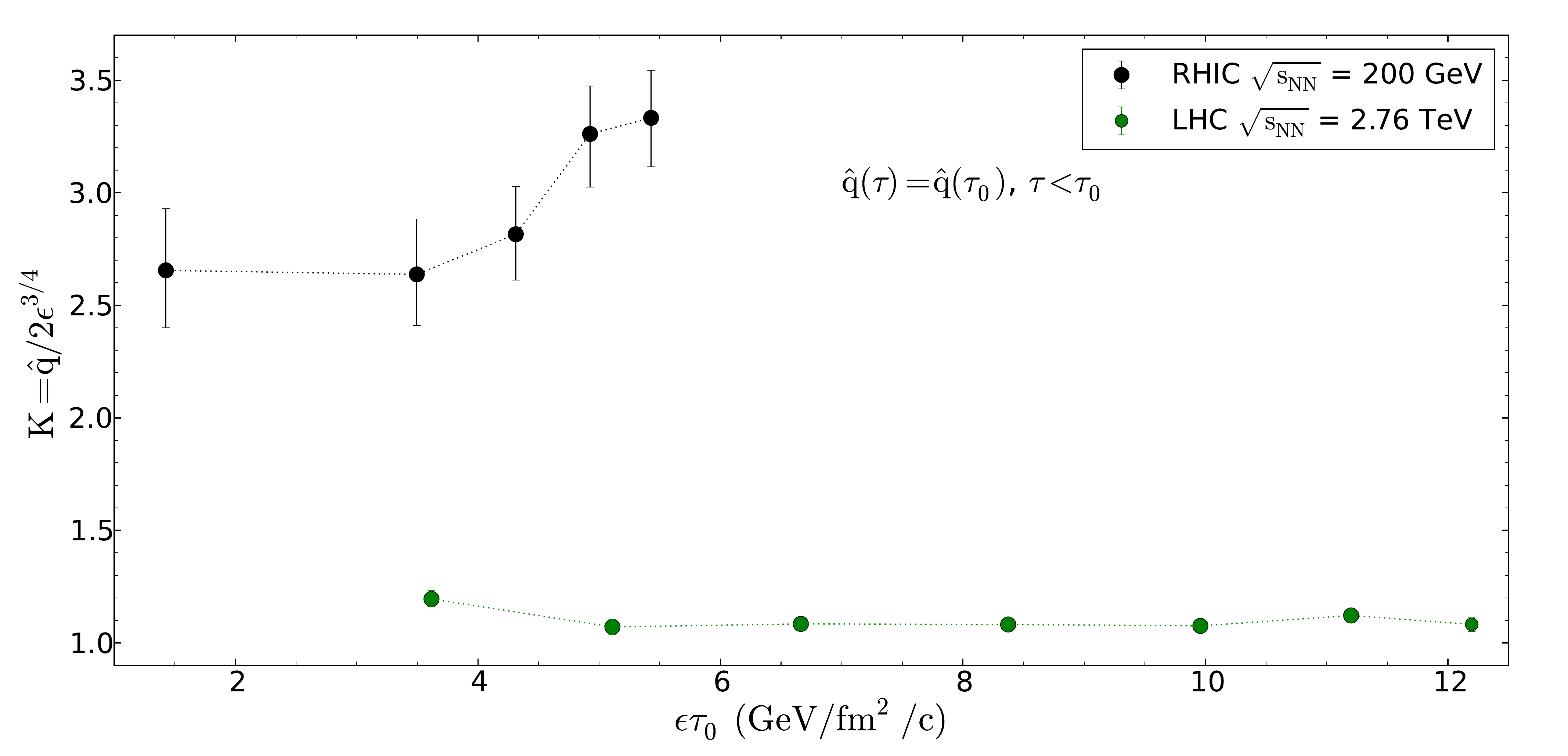}
\caption{$K$-factor at RHIC and LHC energies for different centrality classes versus an estimate of the energy density \cite{Adare:2015bua,Adam:2016thv}}
\label{fig:overlap}
\end{figure*}

There is an overlap on typical temperatures (or energy densities) between semi-peripheral PbPb collisions at the LHC and central Au-Au at RHIC, however, the values of $K$ do not coincide. To illustrate this we show in Figure \ref{fig:overlap}, the $K$-factors obtained for different centralities and energies versus an energy density times formation time $\tau_0$ extracted from the experimental data using Bjorken estimates \cite{Adare:2015bua,Adam:2016thv}.

\section{Conclusions}

We have performed an analysis of the single-inclusive suppression of high-$p_T$ particles as a function of centrality and the energy of the collision. A $K$-factor $\hat q\simeq 2\epsilon^{3/4}$ is defined. This factor is fitted to the corresponding experimental data at RHIC and LHC for different centralities. The fitted value at the LHC is close to unity, while the one at RHIC confirms large corrections to the ideal case. 

The centrality dependences at RHIC and the LHC separately are rather flat. Therefore, the change in the value of $K$ is not only due to the different temperature, as there is a large region of overlap between RHIC and the LHC for different centralities. That is, its value would not depend on local properties of the QGP as temperature, but on global collision variables such as the center of mass energy. This result was completely unexpected.

Our approach has various limitations that may affect the results. First, as we have already mentioned, the quenching weights are based on two assumptions which could fail if color coherence is broken. Multiple soft scattering approximation is used, where the perturbative tails of the distributions are neglected, which may enhance the energy loss. Collisional energy loss is also neglected in our formalism.

\section*{Acknowledgements}
This research was supported by the European Research Council grant HotLHC ERC-2011-StG-279579; the People Programme (Marie Curie Actions) of the European Union's Seventh Framework Programme FP7/2007-2013/ under REA grant agreement \#318921 (NA); Ministerio de Ciencia e Innovaci\'on of Spain under project FPA2014-58293-C2-1-P and FEDER; Xunta de Galicia (Conseller\'{\i}a de Educaci\'on) --- the group is part of the Strategic Unit AGRUP2015/11. C. Andr\'es thanks the Spanish Ministery of Education, Culture and Sports for financial support (grant FPU2013-03558). H.N.\ has received funding from the European Union's Horizon 2020 research and innovation programme under the Marie Sklodowska-Curie grant agreement no.\ 655285."

\bibliographystyle{elsarticle-num}
\bibliography{<your-bib-database>}

\begin{thebibliography}{00}

\bibitem{Baier:2001yt}
  R.~Baier et al.,
  JHEP {\bf 0109} (2001) 033
  doi:10.1088/1126-6708/2001/09/033

\bibitem{Salgado:2002cd}
  C.~A.~Salgado and U.~A.~Wiedemann,
  Phys.\ Rev.\ Lett.\  {\bf 89} (2002) 092303
  doi:10.1103/PhysRevLett.89.092303

\bibitem{Salgado:2003gb}
  C.~A.~Salgado and U.~A.~Wiedemann, 
  Phys.\ Rev.\ D {\bf 68} (2003) 014008
  doi:10.1103/PhysRevD.68.014008


\bibitem{Niemi:2015qia}
      H.~Niemi, K.~J.~Eskola and R.~Paatelainen, 
      Phys.\ Rev.\ C {\bf 93} (2016)  024907
      doi:10.1103/PhysRevC.93.024907

\bibitem{Baier:2002tc}
  R.~Baier,
  Nucl.\ Phys.\ A {\bf 715} (2003) 209
  doi:10.1016/S0375-9474(02)01429-X


\bibitem{Stratmann:2001pb}
  M.~Stratmann and W.~Vogelsang,
  Phys.\ Rev.\ D {\bf 64} (2001) 114007
  doi:10.1103/PhysRevD.64.114007

\bibitem{Nadolsky:2008zw}
  P.~M.~Nadolsky et al.,
  Phys.\ Rev.\ D {\bf 78} (2008) 013004
  doi:10.1103/PhysRevD.78.013004


\bibitem{deFlorian:2007aj}
  D.~de Florian, R.~Sassot and M.~Stratmann,
  Phys.\ Rev.\ D {\bf 75} (2007) 114010
  doi:10.1103/PhysRevD.75.114010
  [hep-ph/0703242 [HEP-PH]];
  Phys.\ Rev.\ D {\bf 76} (2007) 074033
  doi:10.1103/PhysRevD.76.074033 

\bibitem{Eskola:2009uj}
  K.~J.~Eskola, H.~Paukkunen and C.~A.~Salgado,
  JHEP {\bf 0904} (2009) 065
  doi:10.1088/1126-6708/2009/04/065


\bibitem{CasalderreySolana:2012ef}
  J.~Casalderrey-Solana, Y.~Mehtar-Tani, C.~A.~Salgado and K.~Tywoniuk,
  Phys.\ Lett.\ B {\bf 725} (2013) 357
  doi:10.1016/j.physletb.2013.07.046


\bibitem{Blaizot:2012fh}
  J.~P.~Blaizot, F.~Dominguez, E.~Iancu and Y.~Mehtar-Tani,
  JHEP {\bf 1301} (2013) 143
  doi:10.1007/JHEP01(2013)143


\bibitem{Armesto:2007dt}
Nestor Armesto et al.,
JHEP { \bf 02} (2008) 048
doi:10.1088/1126-6708/2008/02/048


\bibitem{Andres:2016iys}
Carlota Andr\'es, N\'estor Armesto, Matthew Luzum, Carlos A.Salgado and P\'ia Zurita,
Eur. Phys. J.C76 (2016) no. 9, 475
doi:10.1140/epjc/s10052-016-4320-5


\bibitem{Adare:2008qa}
  A.~Adare {\it et al.} [PHENIX Collaboration],
  Phys.\ Rev.\ Lett.\  {\bf 101} (2008) 232301
  doi:10.1103/PhysRevLett.101.232301


\bibitem{Abelev:2012hxa}
  B.~Abelev {\it et al.} [ALICE Collaboration],
  Phys.\ Lett.\ B {\bf 720} (2013) 52
  doi:10.1016/j.physletb.2013.01.051


\bibitem{Burke:2013yra}
  K.~M.~Burke {\it et al.}  [JET Collaboration],
  Phys.\ Rev.\ C {\bf 90} (2014) 1,  014909


\bibitem{Adare:2015bua}
  A.~Adare {\it et al.} [PHENIX Collaboration],
  Phys.\ Rev.\ C {\bf 93} (2016) no.2,  024901
  doi:10.1103/PhysRevC.93.024901


\bibitem{Adam:2016thv}
  J.~Adam {\it et al.} [ALICE Collaboration],
  arXiv:1603.04775 [nucl-ex].

\end{thebibliography}

\end{document}